\documentstyle[12pt]{article}
\newcommand{\eqreset}{\setcounter{equation}{0}}
\newtheorem{theorem}{Theorem}[section]

\begin{document}
\title{Analytic Bethe ansatz and functional equations associated with 
 any simple root systems of the Lie superalgebra 
 $sl(r+1|s+1)$}
\author{Zengo Tsuboi
\footnote{This paper was published in 1998 first. 
present address (on December 2009): 
Okayama Institute for Quantum Physics, 
1-9-1 Kyoyama, Okayama 700-0015, Japan} \\ 
Institute of Physics, University of Tokyo \\
 Komaba 3-8-1, Meguro-ku, Tokyo 153 Japan}
\date{}
\maketitle
\begin{abstract}
The Lie superalgebra $sl(r+1|s+1)$ admits several inequivalent choices of 
simple root systems. We have carried out analytic Bethe ansatz for any 
simple root systems of  $sl(r+1|s+1)$. 
We present transfer matrix eigenvalue formulae in dressed vacuum form,
 which are expressed as the  
 Young supertableaux  with
 some semistandard-like conditions. 
 These formulae have determinant expressions,  
which can be viewed as quantum analogue of Jacobi-Trudi and 
Giambelli formulae for $sl(r+1|s+1)$. 
We also propose a class of transfer matrix functional relations,
 which is specialization of Hirota bilinear difference equation. 
 Using the particle-hole transformation, relations among the Bethe 
 ansatz equations for various kinds of simple root systems are discussed.
\end{abstract}
\\\\
Journal-ref: Physica A 252 (1998) 565-585 \\ 
DOI: 10.1016/S0378-4371(97)00625-0
\newpage
%%%%%%%%%%%%%%%%%%%%%%%%%%%%%%%%%%%%%%%%%%%%%
\eqreset
\section{Introduction}
In reference \cite{KS1}, analytic Bethe ansatz \cite{R1,R2}
 was carried out systematically for 
fundamental representations of the Yangians $Y({\cal G})$ \cite{D} 
 associated  
 with classical simple Lie algebras ${\cal G}=B_{r}$, $C_{r}$ and $D_{r}$.
  That is, eigenvalue
 formulas in dressed vacuum form were presented for the 
 commuting transfer matrices of solvable vertex models.
  These formulae are Yangian analogues of the Young tableaux 
 for ${\cal G}$ and obey some semi-standard like 
conditions. It had been proven that they do not have 
poles under the Bethe ansatz equation. Furthermore,
 for ${\cal G}=B_{r}$ case, these
formulae were generalized \cite{KOS}
 to the case of finite dimensional modules 
labeled by skew-Young diagrams $\lambda \subset \mu$.
The eigenvalue formulae of the transfer matrices in dressed vacuum form 
labeled by rectangular Young diagrams $\lambda =\phi, \mu=(m^a)$ 
obey a class of functional relations, the T-system \cite{KNS1} 
(see also, references \cite{BR,KLWZ,K,KP,KS2,LWZ,S2,KNS2}). 
Making use of the T-system,
 we are able to calculate \cite{KNS2} various kinds of physical quantities 
  such as the correlation lengths of the vertex models
  and central charges
 of RSOS models. The T-system is not only a class of 
 transfer matrix functional relations but also a 
 two-dimensional Toda field equation on discrete space time.
 Solving it recursively, we can express its solutions
 in terms of pfaffians or determinants \cite{KOS,KNH,TK,T1}.

In contrast to above mentioned successful story in the T-system 
and the analytic 
 Bethe ansatz for simple Lie algebras, systematic treatment of them 
 for  Lie superalgebras \cite{Ka} had not been studied yet until
 quite recently. 
Studying supersymmetric integrable models is significant not only 
 in mathematical physics but also in condensed matter physics
 (see for example, reference \cite{KE}). 
 For instance, the supersymmetric $t-J$ model received much 
 attention in connection with high $T_{c}$ superconductivity. 
  As is well known, there are several choices of simple root systems for
   a superalgebra. We can construct all the simple root systems, 
    from any one of them by applying repeatedly the reflections with 
    respect to the elements
  of the Weyl supergroup $ {\cal SW}({\cal G})$ \cite{LSS}. 
 The simplest system of simple roots is so
  called distinguished one \cite{Ka}.
 Recently we had executed \cite{T2}
  analytic Bethe ansatz associated mainly with the 
 distinguished simple root system of the Lie superalgebra $sl(r+1|s+1)$ 
 and then established 
 functional relations for commuting family of transfer matrices.
 
 The purpose of this paper is to extend our previous results \cite{T2} 
 to any simple root systems of  ${\cal G}=sl(r+1|s+1)$. 
One can reproduce many of the earlier results \cite{T2}
 if one set the grading 
 parameters (\ref{grading}) to
  $p_{a}=1 : 1\le a \le r+1; p_{a}=-1:r+2 \le a \le r+s+2$.
  Throughout this paper, we often use similar
 notation presented in references \cite{KS1,KOS,TK,T2}.

 We execute analytic Bethe ansatz 
based upon the Bethe ansatz equation (\ref{BAE}) associated 
with any simple root systems of $sl(r+1|s+1)$. 
The observation that the Bathe ansatz equation can be expressed 
by the root system of a Lie algebra is traced back to reference \cite{RW}
 (see also, reference \cite{Kul} for $sl(r+1|s+1)$ case). 
Moreover, Kuniba et.al. \cite{KOS} conjectured that the left hand 
side of the 
 Bethe ansatz equation (\ref{BAE}) can be written as a ratio 
  of some \symbol{96}Drinfeld polynomials' \cite{D}. 
 In addition, extra signs appear in the Bethe ansatz equation.
  This is because in the supersymmetric models, the R-matrix 
  satisfies the graded
Yang-Baxter equation \cite{KulSk} and then 
the transfer matrix is defined as a 
 supertrace of the monodromy matrix.  
 There are several sets of Bethe ansatz equations
  corresponding to the fact that there are several
  choices of simple root systems for a Lie superalgebra. 
  However these sets of the Bethe ansatz equations 
  are connected with each other  
  under the particle-hole transformation. In fact, the eqivalence 
 of these sets of the 
 Bethe ansatz equations was established for $sl(1|2)$ case 
 in references \cite{BCFH,EK} and 
 for $sl(2|2)$ case in reference \cite{EKS2}. 
 Then we discuss relations among these sets of the Bethe 
 ansatz equations for $sl(r+1|s+1)$ 
  and we point out that the particle-hole transformation
  is related with the reflection with respect to the 
  element of the Weyl supergroup for 
 odd simple root $\alpha $ with $(\alpha | \alpha)=0$.

We introduce the Young superdiagram \cite{BB1}. 
 To put it more precisely, this Young superdiagram is different  
 from the classical one in that it carries spectral parameter $u$. 
 In contrast to ordinary Young diagram, 
there is no limitation on the number of rows. We define 
 semi-standard like tableau on it. 
Making use of this tableau, we introduce the function 
${\cal T}_{\lambda \subset \mu}(u)$ (\ref{Tge1}),
 which should be the fusion transfer 
matrix whose auxiliary space is finite dimensional 
 module of super Yangian $Y(sl(r+1|s+1))$ \cite{N} or 
quantum  affine superalgebra $U_{q}(sl(r+1|s+1)^{(1)})$ \cite{Y1,Y2},  
labeled by skew-Young superdiagram $\lambda \subset \mu$.
We can trace the origin of the function ${\cal T}^{1}(u)$ back to  
the eigenvalue formula of 
transfer matrix of the Perk-Schultz model \cite{PS1,PS2,Sc}, 
which is a multi-component generalization of the 
six-vertex model (see also reference \cite{Kul}).  
Furthermore, the function ${\cal T}^{1}(u)$ reduces to 
the eigenvalue formula of transfer matrix derived by 
algebraic Bethe ansatz (For instance, reference \cite{FK}: 
$r=1,s=0$ case; reference \cite{EK}: $r=0,s=1$ case; 
 references \cite{EKS1,EKS2}: $r=s=1$ case).
We prove pole-freeness of ${\cal T}^{a}(u)={\cal T}_{(1^{a})}(u)$,  
essential property in the analytic Bethe ansatz.
Owing to the same mechanism presented in reference \cite{KOS},
 the function
 ${\cal T}_{\lambda \subset \mu}(u)$ has 
determinant expressions whose matrix elements are only 
the functions associated with Young superdiagrams with shape 
$\lambda = \phi $; $\mu =(m)$ or $(1^{a})$. 
They can be viewed as quantum analogue of Jacobi-Trudi and 
Giambelli formulae for  $sl(r+1|s+1)$.  
Then we can easily 
show that the function ${\cal T}_{\lambda \subset \mu}(u)$ 
is free of poles under 
the Bethe ansatz equation (\ref{BAE}).
 We present a class of transfer matrix functional relations 
among the above-mentioned eigenvalue formulae of 
transfer matrix in dressed vacuum form 
associated with rectangular Young superdiagrams. 
It is specialization of Hirota bilinear difference equation \cite{H}, 
which can be proved by the Jacobi identity.

The outline of this paper is given as follows.
In section 2, we brefly review the Lie superalgebra 
${\cal G}=sl(r+1|s+1)$.
In section 3, we execute the analytic Bethe ansatz 
based upon the Bethe ansatz equation (\ref{BAE}) associated 
with any simple root systems. 
We prove pole-freeness of the function 
${\cal T}^{a}(u)={\cal T}_{(1^{a})}(u)$.
In section 4, we propose  functional relations,
 the T-system, associated with the transfer matrces 
 in dressed vacuum form defined in the previous section.
In section 5, using the particle-hole transformation, 
relations among the sets of the Bethe ansatz equations for various 
kinds of simple root systems are discussed. 
Section 6 is devoted to summary and discussion. 

%%%%%%%%%%%%%%%%%%%%%%%%%%%%%%%%%%%%%%%%
\eqreset
\section{The Lie superalgebra $sl(r+1|s+1)$}
\begin{figure}
    \setlength{\unitlength}{0.75pt}
    \begin{center}
    \begin{picture}(380,340) 
      \put(130,20){\circle{20}}
      \put(140,20){\line(1,0){40}}
      \put(190,20){\circle{20}}
      \put(200,20){\line(1,0){40}}
      \put(250,20){\circle{20}}
      \put(182.929,12.9289){\line(1,1){14.14214}}
      \put(182.929,27.07107){\line(1,-1){14.14214}}
      \put(112,0){$\delta_{1}-\delta_{2}$}
      \put(172,0){$\delta_{2}-\epsilon_{1}$}
      \put(232,0){$\epsilon_{1}-\epsilon_{2}$}
      \put(190,75){\vector(0,-1){40}}
      \put(130,100){\circle{20}}
      \put(140,100){\line(1,0){40}}
      \put(190,100){\circle{20}}
      \put(200,100){\line(1,0){40}}
      \put(250,100){\circle{20}}
      \put(122.929,92.9289){\line(1,1){14.14214}}
      \put(122.929,107.07107){\line(1,-1){14.14214}}
      \put(182.929,92.9289){\line(1,1){14.14214}}
      \put(182.929,107.07107){\line(1,-1){14.14214}}
      \put(242.929,92.9289){\line(1,1){14.14214}}
      \put(242.929,107.07107){\line(1,-1){14.14214}}
      \put(112,80){$\delta_{1}-\epsilon_{1}$}
      \put(172,80){$\epsilon_{1}-\delta_{2}$}
      \put(232,80){$\delta_{2}-\epsilon_{2}$}
      \put(160,160){\vector(3,-2){75}}
      \put(220,160){\vector(-3,-2){75}}
      \put(10,180){\circle{20}}
      \put(20,180){\line(1,0){40}}
      \put(70,180){\circle{20}}
      \put(80,180){\line(1,0){40}}
      \put(130,180){\circle{20}}
      \put(122.929,172.9289){\line(1,1){14.14214}}
      \put(122.929,187.07107){\line(1,-1){14.14214}}
      \put(2.929,172.9289){\line(1,1){14.14214}}
      \put(2.929,187.07107){\line(1,-1){14.14214}}
      \put(-8,160){$\delta_{1}-\epsilon_{1}$}
      \put(52,160){$\epsilon_{1}-\epsilon_{2}$}
      \put(112,160){$\epsilon_{2}-\delta_{2}$}
      \put(250,180){\circle{20}}
      \put(260,180){\line(1,0){40}}
      \put(310,180){\circle{20}}
      \put(320,180){\line(1,0){40}}
      \put(370,180){\circle{20}}
      \put(242.929,172.9289){\line(1,1){14.14214}}
      \put(242.929,187.07107){\line(1,-1){14.14214}}
      \put(362.929,172.9289){\line(1,1){14.14214}}
      \put(362.929,187.07107){\line(1,-1){14.14214}}
      \put(232,160){$\epsilon_{1}-\delta_{1}$}
      \put(292,160){$\delta_{1}-\delta_{2}$}
      \put(352,160){$\delta_{2}-\epsilon_{2}$}
      \put(280,240){\vector(3,-2){75}}
      \put(100,240){\vector(-3,-2){75}}
      \put(130,260){\circle{20}}
      \put(140,260){\line(1,0){40}}
      \put(190,260){\circle{20}}
      \put(200,260){\line(1,0){40}}
      \put(250,260){\circle{20}}
      \put(122.929,252.9289){\line(1,1){14.14214}}
      \put(122.929,267.07107){\line(1,-1){14.14214}}
      \put(182.929,252.9289){\line(1,1){14.14214}}
      \put(182.929,267.07107){\line(1,-1){14.14214}}
      \put(242.929,252.9289){\line(1,1){14.14214}}
      \put(242.929,267.07107){\line(1,-1){14.14214}}
      \put(112,240){$\epsilon_{1}-\delta_{1}$}
      \put(172,240){$\delta_{1}-\epsilon_{2}$}
      \put(232,240){$\epsilon_{2}-\delta_{2}$}
      \put(190,315){\vector(0,-1){40}}
      \put(130,340){\circle{20}}
      \put(140,340){\line(1,0){40}}
      \put(190,340){\circle{20}}
      \put(200,340){\line(1,0){40}}
      \put(250,340){\circle{20}}
      \put(182.929,332.9289){\line(1,1){14.14214}}
      \put(182.929,347.07107){\line(1,-1){14.14214}}
      \put(112,320){$\epsilon_{1}-\epsilon_{2}$}
      \put(172,320){$\epsilon_{2}-\delta_{1}$}
      \put(232,320){$\delta_{1}-\delta_{2}$}
  \end{picture}
  \end{center}
  \caption{Dynkin diagrams for the Lie superalgebra $sl(2|2)$:    
A white circle expresses even root; a grey (a cross) 
 circle expresses odd root $\alpha$ with $(\alpha|\alpha)=0$. 
 The topmost Dynkin diagram is the one associated with the 
 distinguished simple root system. 
The odd root $\alpha$ attached to the root of the arrow is 
transfered to the one attached to the arrowhead by 
 the reflection with respect to
 $\omega_{\alpha} \in {\cal SW}({\cal G})$.
%The simple root systems attached to the same Dynkin diagrams are 
%equivalent each other under the Weyl group ${\cal W}({\cal G})$.
}
  \label{dynkin}
\end{figure}
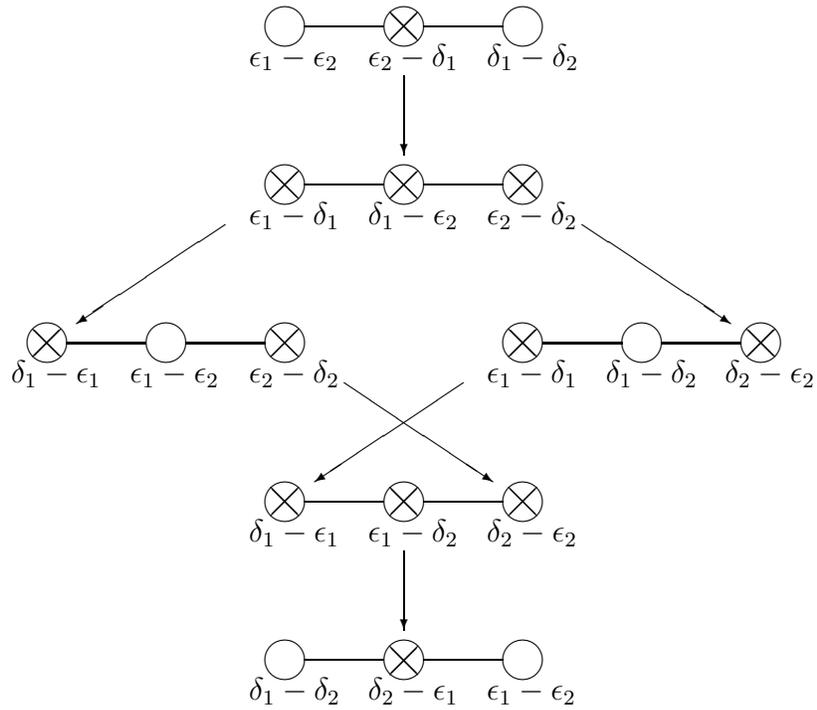

In this section, we brefly review the 
Lie superalgebra ${\cal G}=sl(r+1|s+1)$. 
A Lie superalgebra \cite{Ka} is a ${\bf Z}_2$ graded algebra 
${\cal G} ={\cal G}_{\bar{0}} \oplus {\cal G}_{\bar{1}}$ 
with a product $[\; , \; ]$, whose homogeneous
elements $a\in {\cal G_{\alpha}},b\in {\cal G_{\beta}}$ 
$(\alpha, \beta \in {\bf Z}_2=\{\bar{0},\bar{1} \})$ and
 $c\in {\cal G}$ obey the following relations.
\begin{eqnarray}
\left[a,b\right] & \in & {\cal G}_{\alpha+\beta}, \nonumber \\ 
\left[a,b\right]&=&-(-1)^{\alpha \beta}[b,a], \\
\left[a,[b,c]\right]&=&[[a,b],c]+(-1)^{\alpha \beta} [b,[a,c]].
 \nonumber  
\end{eqnarray}
We can divide the set of non-zero roots into the set of non-zero  
even roots (bosonic roots) $\Delta_0$ and the set of odd roots 
(fermionic roots) $\Delta_1$. For $sl(r+1|s+1)$ case,
 they have the following form
\begin{equation}
\Delta_0=
 \{ \epsilon_{i}-\epsilon_{j} \} \cup \{\delta_{i}-\delta_{j}\}, 
i \ne j ;\quad \Delta_1=\{\pm (\epsilon_{i}-\delta_{j})\}
\end{equation}
where $\epsilon_{1},\dots,\epsilon_{r+1};\delta_{1},\dots,\delta_{s+1}$ 
are basis of dual space of the Cartan subalgebra with the bilinear 
form $(\ |\ )$ such that 
\begin{equation}
 (\epsilon_{i}|\epsilon_{j})=\delta_{i\, j}, \quad 
 (\epsilon_{i}|\delta_{j})=(\delta_{i}|\epsilon_{j})=0 , \quad 
 (\delta_{i}|\delta_{j})=-\delta_{i\, j}. 
\end{equation} 
The Weyl group ${\cal W}({\cal G})$ of a Lie superalgebra ${\cal G}$ is 
generated by the Weyl reflections with respect to the even roots : 
\begin{equation}
\omega_{\alpha}(\beta)=\beta 
     -\frac{2(\alpha | \beta)}{(\alpha | \alpha)} \alpha
\end{equation}
where $\alpha \in \Delta_0, \beta \in 
\Delta_0 \cup \Delta_1 $. 
 Moreover  the Weyl group ${\cal W}({\cal G})$ can be
  extended to the Weyl supergroup ${\cal SW}({\cal G})$ \cite{LSS} by 
  adding the reflections with respect to the odd roots: 
\begin{equation}
\omega_{\alpha}(\beta)=
\left\{
  \begin{array}{@{\,}ll}
   \beta -\frac{2(\alpha | \beta)}{(\alpha | \alpha)} \alpha & 
   \mbox{for} \quad (\alpha | \alpha) \ne 0 \\ 
   \beta +\alpha  &  \mbox{for} \quad (\alpha | \alpha) = 0 \quad 
   \mbox{and} \quad (\alpha | \beta) \ne 0 \\ 
   \beta &  \mbox{for} \quad (\alpha | \alpha)=(\alpha | \beta) =0 \\
   -\alpha & \mbox{for} \quad  \beta=\alpha  
  \end{array}
\right. 
 \end{equation}
where  $\alpha \in \Delta_1, \beta \in 
\Delta_0 \cup \Delta_1 $. 
Note that $\Delta_0$ and $\Delta_1 $ are invariant under 
$\omega_{\alpha} \in {\cal W}({\cal G})$; are not invariant under 
$\omega_{\alpha} \in {\cal SW}({\cal G})$ with $(\alpha | \alpha)=0$. 
There are several choices of simple root systems depending on 
choices of Borel subalgebras.
The simplest system of simple roots is so called
 distinguished one \cite{Ka}.
For example,  
the distinguished simple root system
 $\{\alpha_1,\dots,\alpha_{r+s+1} \}$
 of $sl(r+1|s+1)$ has the form 
 \begin{eqnarray}
   &&\alpha_i = \epsilon_{i}-\epsilon_{i+1}
    \quad i=1,2,\dots,r, \nonumber \\
   &&\alpha_{r+1} = \epsilon_{r+1}-\delta_{1}  \\ 
   && \alpha_{j+r+1} = \delta_{j}-\delta_{j+1} ,
    \quad j=1,2,\dots,s \nonumber   
 \end{eqnarray}
where $\{\alpha_i \}_{i \ne r+1}$ are even roots and $\alpha_{r+1}$ 
is an odd root with $(\alpha_{r+1} | \alpha_{r+1})=0$. 
One can construct all the simple root systems, unequivalent with respect 
to ${\cal W}({\cal G})$, from any one of them by applying repeatedly
 the reflections with respect to
  $\omega_{\alpha} \in {\cal SW}({\cal G})$
 with $(\alpha | \alpha )=0$ (see, figure \ref{dynkin}).
We define the sets 
\begin{eqnarray}
      J=\{ 1,2,\dots,r+s+2\} 
  \label{set}
\end{eqnarray}
with the total order 
\begin{eqnarray} 
 1\prec 2 \prec \cdots \prec r+s+2 . \label{order}
\end{eqnarray}
Divide the set $J$ into two disjoint sets 
\begin{eqnarray}
  J=J_{+} \bigcup J_{-}, & \qquad & 
  J_{+} \bigcap J_{-} = \phi, \nonumber \\
  J_{+}=\{ i_{1},i_{2},\dots,i_{r+1}\} , &\quad &   
  J_{-}=\{ j_{1},j_{2},\dots,j_{s+1}\}
  \label{disj}   
\end{eqnarray}
with the ordering 
\begin{eqnarray} 
 i_{1} \prec i_{2} \prec \cdots \prec i_{r+1},\quad 
 j_{1} \prec j_{2} \prec \cdots \prec j_{s+1} . 
\end{eqnarray}
For any element of $J$, we introduce the grading 
\begin{equation}
      p_{a}=\left\{
              \begin{array}{@{\,}ll}
                1  & \mbox{for $a \in J_{+}$}  \\ 
                -1 & \mbox{for $a \in J_{-}$ }
                 \quad .
              \end{array}
            \right. \label{grading}
\end{equation}
Using this grading parameters 
$\{p_{j} \}$, one can express Cartan matrix as follows 
\begin{equation}
(\alpha_{k}|\alpha_{l})=(p_{k}+p_{k+1})\delta_{k\>l}
-p_{k+1}\delta_{k+1\> l}
-p_{k}\delta_{k\> l+1}. \label{cartangr}
\end{equation}
%%%%%%%%%%%%%%%%%%%%%%%%%%%%%%%%%%%%%%%%%%%%%
\eqreset
\section{Analytic Bethe ansatz}
Consider the following type of the Bethe
 ansatz equation (cf. references \cite{Kul,RW,KOS,Sc}).
\begin{eqnarray}
 -\frac{P_{a}(u_k^{(a)}+\zeta_{a})}
 {P_{a}(u_k^{(a)}-\zeta_{a})}
   &=&(-1)^{{\rm deg}(\alpha_a)} 
    \prod_{b=1}^{r+s+1}\frac{Q_{b}(u_k^{(a)}+(\alpha_a|\alpha_b))}
           {Q_{b}(u_k^{(a)}-(\alpha_a|\alpha_b))}, \label{BAE} \\ 
        Q_{a}(u)&=& \prod_{j=1}^{N_{a}}[u-u_j^{(a)}],
        \label{Q_a} \\ 
        P_{a}(u)&=& \prod_{j=1}^{N}P_{a}^{(j)}(u), \\  
        P_{a}^{(j)}(u)&=&[u-w_{j}]^{\delta_{a,1}} \label{drinfeld}
\end{eqnarray}
where $[u]=(q^u-q^{-u})/(q-q^{-1})$; $N_{a} \in {\bf Z }_{\ge 0}$; 
$u, w_{j}\in {\bf C}$; $a,k \in {\bf Z}$ 
($1\le a \le r+s+1$,$\ 1\le k\le N_{a}$), $\zeta_{1}=p_{1}$  
  and  
\begin{eqnarray}
    {\rm deg}(\alpha_a)&=&\left\{
              \begin{array}{@{\,}ll}
                0  & \mbox{for even root} \\ 
                1 & \mbox{for odd root} 
              \end{array}
              \right. \\ 
            &=& \frac{1-p_{a}p_{a+1}}{2} . \nonumber 
\end{eqnarray}
Particularly for distinguished simple root of 
 $sl(r+1|s+1)$, we have ${\rm deg}(\alpha_{a})=\delta_{a,r+1}$.
In the present paper, we suppose that $q$ is generic.
The left hand side of the Bethe ansatz equation (\ref{BAE}) 
is connected with the quantum space.
We suppose that it is the ratio of some 
\symbol{96}Drinfeld polynomials'  labeled by skew-Young diagrams
  $\tilde{\lambda} \subset \tilde{\mu}$ (cf. reference \cite{KOS}).
  For simplicity, we deal only with  the case 
   $\tilde{\lambda}=\phi, \tilde{\mu}=(1) $.
The generalization to the case for any skew-Young diagram
will be accomplished by the empirical procedures 
given in reference \cite{KOS}.
 The factor $(-1)^{{\rm deg}(\alpha_a)}$ of 
the Bethe ansatz equation (\ref{BAE}) exists so as to make
 the transfer matrix to be the supertrace of the monodromy matrix. 
 Note that the Bethe ansatz equation (\ref{BAE}) is invariant 
 under the Weyl group ${\cal W}({\cal G})$ since 
 $(\omega_{\beta}(\alpha)|\omega_{\beta}(\gamma))=
 (\alpha|\gamma)$ for $\alpha,\gamma \in \Delta_{0} \cup \Delta_{1}$ and 
 $\beta \in \Delta_{0}$.
 
For any $a \in J $, set
\begin{eqnarray}
   z(a;u)=\psi_{a}(u)
     \frac{Q_{a-1}(u+\sum_{j=1}^{a-1}p_{j}+2p_{a})
     Q_{a}(u+\sum_{j=1}^{a}p_{j}-2p_{a})}
     {Q_{a-1}(u+\sum_{j=1}^{a-1}p_{j})Q_{a}(u+\sum_{j=1}^{a}p_{j})} 
\end{eqnarray}
where $Q_{0}(u)=1, Q_{r+s+2}(u)=1$ and 
\begin{equation}
  \psi_{a}(u)=
   \left\{
    \begin{array}{@{\,}ll}
      P_{1}(u+2p_{1}) & \mbox{for } \quad a=1 \\ 
      P_{1}(u) & \mbox{for } \quad a \in J-\{1\}
    \end{array} \label{psi}
   \right. .
\end{equation}
In the present paper, we frequently express the function $z(a;u)$ as the 
box $\framebox{a}_{u}$, whose spectral parameter $u$ will often 
 be abbreviated. Under the Bethe ansatz equation (\ref{BAE}), 
 the following relations are valid 
\begin{eqnarray}
 Res_{u=-\sum_{j=1}^{b}p_{j}+u_{k}^{(b)}}
    (p_{b}z(b;u)+p_{b+1}z(b+1;u))=0 , \quad 
     b \in J-\{r+s+2 \} . \label{res1} 	
\end{eqnarray}
It was pointed out \cite{KS1} that the dressed vacuum form in the analytic 
Bethe ansatz for fundamental representations of Yangians 
$Y({\cal G})$ associated with simple 
 Lie algebras ${\cal G}=B_{r},C_{r},D_{r}$ have similar 
 structure of the crystal graph 
  \cite{KN,Na}. This is also the case with 
  ${\cal G}=sl(r+1|s+1)$. Actually, 
we can express the relation ($\ref{res1}$) schematically as follows:  
\begin{equation}
 p_{1} \framebox{1} \stackrel{1}{\longrightarrow} 
 p_{2} \framebox{2} \stackrel{2}{\longrightarrow} 
 \cdots \stackrel{r+s+1}{\longrightarrow} p_{r+s+2}  \framebox{r+s+2}
 \label{cg} 
\end{equation}
where the number $b$ on the arrow represents 
the \symbol{96}color\symbol{39} (superscript of $u_{k}^{(b)}$) of 
 the common pole $-\sum_{j=1}^{b}p_{j}+u_{k}^{(b)}$ 
 of the functions $z(b;u)$ and $z(b+1;u)$.

We will use the functions ${\cal T}^{a}(u)$ 
and ${\cal T}_{m}(u)$ 
 ($a,m \in {\bf Z }$; $u \in {\bf C }$) 
 determined by the non-commutative generating series of the form  
\begin{eqnarray}
 & &   (1+z(r+s+2;u)X)^{p_{r+s+2}}\cdots (1+z(r+2;u)X)^{p_{r+2}}
  \nonumber \\
 &&\times  (1+z(r+1;u)X)^{p_{r+1}}\cdots (1+z(1;u)X)^{p_{1}}
   \nonumber \\        
    &&=\sum_{a=-\infty}^{\infty} 
         {\cal T}^{a}(u+a-1)X^{a},
         \label{generating}
\end{eqnarray} 
\begin{eqnarray}
&& (1-z(1;u)X)^{-p_{1}}\cdots (1-z(r+1;u)X)^{-p_{r+1}} \nonumber \\
 &&\times  (1-z(r+2;u)X)^{-p_{r+2}}\cdots (1-z(r+s+2;u)X)^{-p_{r+s+2}}
 \nonumber \\
 &&= \sum_{m=-\infty}^{\infty} {\cal T}_{m}(u+m-1)X^{m} \label{generating2}
\end{eqnarray} 
where $X$ is a shift operator $X=e^{2\partial_{u}}$. 
In particular,  we have ${\cal T}^{0}(u)=1$; 
${\cal T}_{0}(u)=1$; ${\cal T}^{a}(u)=0$ for $a<0$; 
 ${\cal T}_{m}(u)=0$ for $m<0$. 
We note that the origin of the functions 
${\cal T}^{1}(u),{\cal T}_{1}(u)$ and 
the Bethe ansatz equation (\ref{BAE}) with (\ref{cartangr}) 
trace back to the eigenvalue formula of transfer matrix and the Bethe 
ansatz equation for the Perk-Schultz model \cite{Sc} but the vacuum part,
 some gauge factors and extra signs after some redefinition. 
 (See also, reference \cite{Kul}). 
 As for the relation between fundamental $L$ operator
  and the function ${\cal T}^{1}(u)$, 
  see, for example, Appendix A in reference \cite{T2}.

Let $\lambda \subset \mu$ be a skew-Young superdiagram labeled by 
the sequences of non-negative integers 
$\lambda =(\lambda_{1},\lambda_{2},\dots)$ and 
$\mu =(\mu_{1},\mu_{2},\dots)$ such that
$\mu_{i} \ge \lambda_{i}: i=1,2,\dots;$  
$\lambda_{1} \ge \lambda_{2} \ge \dots \ge 0$;  
$\mu_{1} \ge \mu_{2} \ge \dots \ge 0$ and 
$\lambda^{\prime}=(\lambda_{1}^{\prime},\lambda_{2}^{\prime},\dots)$ 
be the conjugate of $\lambda $.
%
%\begin{figure}
%  \begin{center}
%    \setlength{\unitlength}{2pt}
%    \begin{picture}(50,55) 
%      \put(0,0){\line(0,1){20}}
%      \put(10,0){\line(0,1){30}}
%      \put(20,20){\line(0,1){20}}
%      \put(30,20){\line(0,1){20}}
%      \put(40,20){\line(0,1){30}}
%      \put(50,30){\line(0,1){20}}
%      \put(0,0){\line(1,0){10}}      
%      \put(0,10){\line(1,0){10}}
%      \put(0,20){\line(1,0){40}}
%      \put(10,30){\line(1,0){40}}
%      \put(20,40){\line(1,0){30}}
%      \put(40,50){\line(1,0){10}}
%    \end{picture}
%  \end{center}
%  \caption{Young superdiagram with shape $\lambda \subset \mu$ : 
%  $\lambda=(4,2,1,0,0)$, $\mu=(5,5,4,1,1)$}
%  \label{young}
%\end{figure}
%
%\begin{figure}
%  \begin{center}
%    \setlength{\unitlength}{2pt}
%    \begin{picture}(50,55) 
%      \put(0,0){\line(0,1){10}}
%      \put(10,0){\line(0,1){30}}
%      \put(20,0){\line(0,1){40}}
%      \put(30,10){\line(0,1){40}}
%      \put(40,40){\line(0,1){10}}
%      \put(50,40){\line(0,1){10}}
%      \put(0,0){\line(1,0){20}}      
%      \put(0,10){\line(1,0){30}}
%      \put(10,20){\line(1,0){20}}
%      \put(10,30){\line(1,0){20}}
%      \put(20,40){\line(1,0){30}}
%      \put(30,50){\line(1,0){20}}
%    \end{picture}
%  \end{center}
%  \caption{Young superdiagram with shape $\lambda ^{\prime} \subset
%   \mu^{\prime}$ : $\lambda^{\prime}=(3,2,1,1,0)$,
%    $\mu^{\prime}=(5,3,3,3,2)$}
%    \label{conjyoung}
%%\end{figure}
%
We assign a coordinates $(i,j)\in {\bf Z}^{2}$ 
on this skew-Young  superdiagram $\lambda \subset \mu$ 
such that the row index $i$ increases as we go downwords and the column 
index $j$ increases as we go from left to right and that 
$(1,1)$ is on the top left corner of $\mu$.
We define an admissiable tableau $b$ 
on the skew-Young superdiagram 
$\lambda \subset \mu$ as a set of element $b(i,j)\in J$ 
 labeled by the coordinates 
$(i,j)$ mentioned above, with the following rule 
(admissibility conditions).
\begin{enumerate}
\item 
For any elements of $J$,
\begin{equation}
 b(i,j) \preceq b(i,j+1),\quad b(i,j) \preceq b(i+1,j).
\end{equation}
\item 
For any elements of $J_{+}$,
\begin{equation}
 b(i,j) \prec b(i+1,j).
\end{equation}
\item 
For any elements of $J_{-}$,
\begin{equation}
 b(i,j) \prec b(i,j+1).
\end{equation}
\end{enumerate}
Let $B(\lambda \subset \mu)$ be 
the set of admissible tableaux 
 on $\lambda \subset \mu$.
For any skew-Young superdiagram $\lambda \subset \mu$, 
define the function ${\cal T}_{\lambda \subset \mu}(u)$ as follows
\begin{equation}
 {\cal T}_{\lambda \subset \mu}(u)=
\sum_{b \in B(\lambda \subset \mu)}
\prod_{(i,j) \in (\lambda \subset \mu)}
p_{b(i,j)}
z(b(i,j);u-\mu_{1}+\mu_{1}^{\prime}-2i+2j)	
\label{Tge1}
\end{equation}
where the product is taken over the coordinates $(i,j)$ on
 $\lambda \subset \mu$.
%
%Especially, for an enpty diagram 
%$\phi$, set ${\cal T}_{\phi}(u)=1$. 
The following relations should be
 valid by the same reason mentioned in \cite{KOS}, 
that is, they will be verified by induction on $\mu_{1}$ 
or $\mu_{1}^{\prime}$.
\begin{eqnarray}
 {\cal T}_{\lambda \subset \mu}(u)&=&{\rm det}_{1 \le i,j \le \mu_{1}}
    ({\cal T}^{\mu_{i}^{\prime}-\lambda_{j}^{\prime}-i+j}
    (u-\mu_{1}+\mu_{1}^{\prime}-\mu_{i}^{\prime}-\lambda_{j}^{\prime}+i+j-1))	
	\label{Jacobi-Trudi1} \\ 
	 &=&{\rm det}_{1 \le i,j \le \mu_{1}^{\prime}}
    ({\cal T}_{\mu_{j}-\lambda_{i}+i-j}
    (u-\mu_{1}+\mu_{1}^{\prime}+\mu_{j}+\lambda_{i}-i-j+1))	.
	\label{Jacobi-Trudi2} 
\end{eqnarray}
For instance, for $sl(2|1)$: $\lambda=\phi; \mu=(2^{1},1^{1}); 
J_{+}=\{1,3\}; J_{-}=\{2\}$ case, we obtain  
\begin{eqnarray}
{\cal T}_{(2^{1},1^{1})}(u) &=&
  -\> \begin{array}{|c|c|}\hline 
     	1 & 1 \\ \hline
     	2   \\ \cline{1-1} 
     \end{array}
   +\begin{array}{|c|c|}\hline 
     	1 & 1 \\ \hline
     	3   \\ \cline{1-1} 
     \end{array}
  +
    \begin{array}{|c|c|}\hline 
     	1 & 2 \\ \hline
     	2   \\ \cline{1-1} 
     \end{array}
   -\begin{array}{|c|c|}\hline 
     	1 & 2 \\ \hline
     	3   \\ \cline{1-1} 
     \end{array} \nonumber \\  
   &-&\begin{array}{|c|c|}\hline 
     	1 & 3 \\ \hline
     	2   \\ \cline{1-1} 
     \end{array}
   +\begin{array}{|c|c|}\hline 
     	1 & 3 \\ \hline
     	3   \\ \cline{1-1} 
     \end{array}
  +
    \begin{array}{|c|c|}\hline 
     	2 & 3 \\ \hline
     	2   \\ \cline{1-1} 
     \end{array}
   -\begin{array}{|c|c|}\hline 
     	2 & 3 \\ \hline
     	3   \\ \cline{1-1} 
     \end{array}   \nonumber \\
    & =&
    P_1(u-2)P_1(u+2)
\Bigl\{
-P_1(u+4)\frac{Q_1(u-3)Q_2(u)}{Q_1(u+3)Q_2(u-2)} \nonumber \\
&+&P_1(u+4)  \frac{Q_1(u-1)Q_2(u)}{Q_1(u+3)Q_2(u-2)}  \nonumber \\ 
&+& P_1(u+2)\frac{Q_1(u-3)Q_2(u)Q_2(u+4)}
               {Q_1(u+3)Q_2(u-2)Q_2(u+2)} \nonumber \\ 
&-& P_1(u+2)\frac{Q_1(u-1)Q_2(u)Q_2(u+4)}{Q_1(u+3)Q_2(u-2)Q_2(u+2)}
 \nonumber \\
&-&P_1(u+2)\frac{Q_1(u-3)Q_2(u)Q_2(u+4)}
              {Q_1(u+1)Q_2(u-2)Q_2(u+2)} \nonumber \\
&+&P_1(u+2)  \frac{Q_1(u-1)Q_2(u)Q_2(u+4)}{Q_1(u+1)Q_2(u-2)Q_2(u+2)}
 \nonumber \\  
&+& P_1(u)\frac{Q_1(u-3)Q_2(u+4)}
               {Q_1(u+1)Q_2(u-2)} \nonumber \\ 
&-& P_1(u)\frac{Q_1(u-1)Q_2(u+4)}{Q_1(u+1)Q_2(u-2)} 
\Bigr\}  \label{example} \\ 
&=&\begin{array}{|cc|} 
     	{\cal T}^{2}(u-1) & {\cal T}^{3}(u) \\
     	     1            & {\cal T}^{1}(u+2) \\  
     \end{array}
     \nonumber 
 \end{eqnarray}
 where 
\begin{eqnarray}
{\cal T}^{1}(u)
&=&\begin{array}{|c|}\hline 
     	1  \\ \hline 
     \end{array}
-\begin{array}{|c|}\hline 
     	2  \\ \hline 
     \end{array}
+\begin{array}{|c|}\hline 
     	3 \\ \hline 
     \end{array} \nonumber \\ 
&=&
P_1(u+2)\frac{Q_1(u-1)}{Q_1(u+1)}
-P_1(u)\frac{Q_1(u-1)Q_2(u+2)}{Q_1(u+1)Q_2(u)} \nonumber \\
&+&P_1(u)\frac{Q_2(u+2)}{Q_2(u)},  \\ 
{\cal T}^{2}(u)
&=& 
 -\>  \begin{array}{|c|}\hline 
     	1  \\ \hline 
     	2 \\ \hline
     \end{array}
+   \begin{array}{|c|}\hline 
     	1  \\ \hline 
     	3 \\ \hline
     \end{array}
+    \begin{array}{|c|}\hline 
     	2  \\ \hline 
     	2 \\ \hline
     \end{array}
-    \begin{array}{|c|}\hline 
     	2  \\ \hline 
     	3 \\ \hline
     \end{array} \nonumber \\
&=& P_1(u-1) 
\Bigl\{
 -P_1(u+3)\frac{Q_1(u-2)Q_2(u+1)}{Q_1(u+2)Q_2(u-1)} \nonumber \\ 
&+&P_1(u+3)\frac{Q_1(u)Q_2(u+1)}{Q_1(u+2)Q_2(u-1)}  \\ 
&+&P_1(u+1)\frac{Q_1(u-2)Q_2(u+3)}{Q_1(u+2)Q_2(u-1)}
-P_1(u+1)\frac{Q_1(u)Q_2(u+3)}{Q_1(u+2)Q_2(u-1)}
\Bigr\}, \nonumber \\
{\cal T}^{3}(u)
&=& 
    \begin{array}{|c|}\hline 
     	1  \\ \hline 
     	2 \\ \hline
     	2 \\ \hline
     \end{array}
-   \begin{array}{|c|}\hline 
     	1  \\ \hline 
     	2 \\ \hline
     	3 \\ \hline
     \end{array}
-    \begin{array}{|c|}\hline 
     	2  \\ \hline 
     	2 \\ \hline
     	2 \\ \hline
     \end{array}
+    \begin{array}{|c|}\hline 
     	2  \\ \hline 
     	2 \\ \hline
     	3 \\ \hline
     \end{array} 
     \nonumber \\ 
&=&P_1(u-2)P_1(u) 
\Bigl\{
 P_1(u+4)\frac{Q_1(u-3)Q_2(u+2)}{Q_1(u+3)Q_2(u-2)} \nonumber \\ 
&-&P_1(u+4)\frac{Q_1(u-1)Q_2(u+2)}{Q_1(u+3)Q_2(u-2)}  \\ 
&-&P_1(u+2)\frac{Q_1(u-3)Q_2(u+4)}{Q_1(u+3)Q_2(u-2)}
+P_1(u+2)\frac{Q_1(u-1)Q_2(u+4)}{Q_1(u+3)Q_2(u-2)}
\Bigr\} \nonumber \\ 
&=& -\> \frac{{\cal T}_{(2^2)}(u)}{P_{1}(u+2)}. \nonumber  
 \end{eqnarray}
Remark1: If we drop the $u$ dependence of (\ref{Jacobi-Trudi1}) and 
(\ref{Jacobi-Trudi2}) for $p_{1}=p_{2}=\cdots =p_{r+1}=1,
p_{r+2}=p_{r+3}=\cdots =p_{r+s+2}=-1$, they reduce to classical 
Jacobi-Trudi and Giambelli formulae for $sl(r+1|s+1)$ \cite{BB1,PT}, 
which give  us classical (super) characters. 
This fact confirms \symbol{96}character limit\symbol{39} \cite{KS1}  
of the eigenvalue formula for the transfer matrix. \\ 
Remark2: In the case $\lambda =\phi$ and $s=-1$, $(\ref{Jacobi-Trudi1})$
and $(\ref{Jacobi-Trudi2})$  
reduce to the quantum analogue of Jacobi-Trudi and Giambelli 
formulae for $sl_{r+1}$ presented in reference \cite{BR}.\\
Remark3:  $(\ref{Jacobi-Trudi1})$ and $(\ref{Jacobi-Trudi2})$ 
have the same form as 
the quantum Jacobi-Trudi and Giambelli formulae for 
$U_{q}(B_{n}^{(1)})$ in reference \cite{KOS}, but the function 
${\cal T}^{a}(u)$ is quite different as we can be easily seen from 
(\ref{generating}) and (\ref{generating2}). 

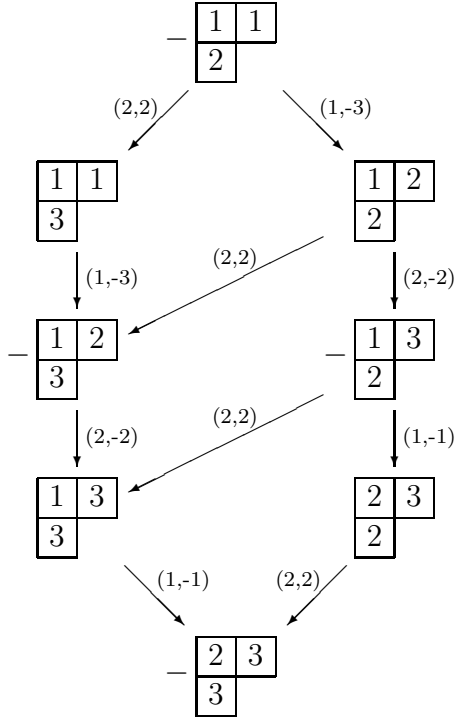
\begin{figure}
    \setlength{\unitlength}{1.5pt}
    \begin{center}
    \begin{picture}(100,180) 
      \put(40,0){\line(1,0){10}}
      \put(40,10){\line(1,0){20}}
      \put(40,20){\line(1,0){20}}
      \put(40,0){\line(0,1){20}}
      \put(50,0){\line(0,1){20}}
      \put(60,10){\line(0,1){10}}
      \put(22,38){\vector(1,-1){15}}
      \put(30,33){\scriptsize{(1,-1)}}
      \put(78,38){\vector(-1,-1){15}}
      \put(60,33){\scriptsize{(2,2)}}
      \put(32,9){$-$}
      \put(43,3){$3$}
      \put(43,13){$2$}
      \put(53,13){$3$}
      \put(0,40){\line(1,0){10}}
      \put(0,50){\line(1,0){20}}
      \put(0,60){\line(1,0){20}}
      \put(0,40){\line(0,1){20}}
      \put(10,40){\line(0,1){20}}
      \put(20,50){\line(0,1){10}}
      \put(10,77){\vector(0,-1){14}}
      \put(12,69){\scriptsize{(2,-2)}}
      \put(3,43){$3$}
      \put(3,53){$1$}
      \put(13,53){$3$}
      \put(80,40){\line(1,0){10}}
      \put(80,50){\line(1,0){20}}
      \put(80,60){\line(1,0){20}}
      \put(80,40){\line(0,1){20}}
      \put(90,40){\line(0,1){20}}
      \put(100,50){\line(0,1){10}}
      \put(90,77){\vector(0,-1){14}}
      \put(92,69){\scriptsize{(1,-1)}}
      \put(83,43){$2$}
      \put(83,53){$2$}
      \put(93,53){$3$}
      \put(0,80){\line(1,0){10}}
      \put(0,90){\line(1,0){20}}
      \put(0,100){\line(1,0){20}}
      \put(0,80){\line(0,1){20}}
      \put(10,80){\line(0,1){20}}
      \put(20,90){\line(0,1){10}}
      \put(10,117){\vector(0,-1){14}}
      \put(12,109){\scriptsize{(1,-3)}}
      \put(-8,89){$-$}
      \put(3,83){$3$}
      \put(3,93){$1$}
      \put(13,93){$2$}
      \put(80,80){\line(1,0){10}}
      \put(80,90){\line(1,0){20}}
      \put(80,100){\line(1,0){20}}
      \put(80,80){\line(0,1){20}}
      \put(90,80){\line(0,1){20}}
      \put(100,90){\line(0,1){10}}
      \put(90,117){\vector(0,-1){14}}
      \put(92,109){\scriptsize{(2,-2)}}
      \put(72,89){$-$}
      \put(83,83){$2$}
      \put(83,93){$1$}
      \put(93,93){$3$}
      \put(0,120){\line(1,0){10}}
      \put(0,130){\line(1,0){20}}
      \put(0,140){\line(1,0){20}}
      \put(0,120){\line(0,1){20}}
      \put(10,120){\line(0,1){20}}
      \put(20,130){\line(0,1){10}}
      \put(38,158){\vector(-1,-1){15}}
      \put(19,152){\scriptsize{(2,2)}}
      \put(3,123){$3$}
      \put(3,133){$1$}
      \put(13,133){$1$}
      \put(80,120){\line(1,0){10}}
      \put(80,130){\line(1,0){20}}
      \put(80,140){\line(1,0){20}}
      \put(80,120){\line(0,1){20}}
      \put(90,120){\line(0,1){20}}
      \put(100,130){\line(0,1){10}}
      \put(62,158){\vector(1,-1){15}}
      \put(71,152){\scriptsize{(1,-3)}}
      \put(83,123){$2$}
      \put(83,133){$1$}
      \put(93,133){$2$}
      \put(40,160){\line(1,0){10}}
      \put(40,170){\line(1,0){20}}
      \put(40,180){\line(1,0){20}}
      \put(40,160){\line(0,1){20}}
      \put(50,160){\line(0,1){20}}
      \put(60,170){\line(0,1){10}}
      \put(32,169){$-$}
      \put(43,163){$2$}
      \put(43,173){$1$}
      \put(53,173){$1$}
      \put(72,81){\vector(-2,-1){49}}
      \put(44,74){\scriptsize{(2,2)}}
      \put(72,121){\vector(-2,-1){49}}
      \put(44,114){\scriptsize{(2,2)}}
  \end{picture}
  \end{center}
  \caption{The \symbol{96}Bethe-strap\symbol{39} structure of the 
  function ${\cal T}_{(2^{1},1^{1})}(u)$  for
   the Lie superalgebra $sl(2|1)$ with the grading 
   $p_{1}=1,p_{2}=-1,p_{3}=1$:  
 The pair $(a,b)$ denotes the common pole $u_{k}^{(a)}+b$ of the pair   
 of the tableaux connected by the arrow.   
 This common pole vanishes under the Bethe ansatz equation.
 The topmost tableau coresponds to the 
 \symbol{96}hightest weight vector\symbol{39}, 
 which is called the \symbol{96}top term\symbol{39}.}
  \label{best}
\end{figure}

The following Theorem is a generalization of the Theorem 
in reference \cite{T2}.
We will present a detailed proof here partly because it 
is essential in the analytic Bethe ansatz and partly because for 
reader's convenience. 
\begin{theorem}\label{polefree}
For any integer $a$, the function ${\cal T}^a(u)$  
is free of poles under the condition that
the Bethe ansatz equation {\rm (\ref{BAE})} is valid.   
\end{theorem}
Proof.  
For simplicity, we assume that the vacuum parts are formally trivial, 
that is, the left hand side of the 
 Bethe ansatz equation (\ref{BAE}) is constantly $-1$. 
We prove that ${\cal T}^a(u)$ is free of color $b$ pole, namely, 
$Res_{u=u_{j}^{(b)}+\cdots}{\cal T}^a(u)=0$ for any $b \in J-\{r+s+2 \}$
 under the condition that the Bethe
  ansatz equation (\ref{BAE}) is valid. 
 The function $z(c;u)=\framebox{$c$}_{u}$ with $c\in J $ has 
the color $b$ pole only for $c=b$ or $b+1$, so we shall trace only 
\framebox{$b$} or \framebox{$b+1$}.
Denote $S_{k}$ the partial sum of ${\cal T}^a(u)$, which contains 
k boxes among \framebox{$b$} or \framebox{$b+1$}.
 Apparently, $S_{0}$ does not have 
color $b$ pole. Now we examine $S_{1}$, which is the summation 
 of the tableaux of the following form 
\begin{equation} 
\begin{array}{|c|}\hline
    \xi \\ \hline 
    b   \\ \hline 
   \zeta \\ \hline
\end{array}
\qquad \qquad 
\begin{array}{|c|}\hline
    \xi \\ \hline 
    b+1 \\ \hline 
   \zeta \\ \hline
\end{array} \label{tableaux1}
\end{equation}
where \framebox{$\xi$} and \framebox{$\zeta$} are columns whose 
total length are $a-1$ and they do not involve 
 \framebox{$b$} and \framebox{$b+1$}.
 Thanks to the relation (\ref{res1}), 
color $b$ residues in these tableaux (\ref{tableaux1}) cancel each 
other under the Bethe ansatz equation (\ref{BAE}). Then we deal with 
$S_{k}$ only for $2 \le k \le a $ from now on. \\
The case $ b, b+1 \in J_{+}$ : In this case, only the case for 
$k=2$ should be considered because 
\framebox{$b$} or \framebox{$b+1$} appear at 
 most twice in one column. $S_{2}$ is the 
summation of the tableaux of the following form  
\begin{equation}
\begin{array}{|c|l}\cline{1-1}
    \xi & \\ \cline{1-1} 
    b   & _v  \\ \cline{1-1} 
    b+1 & _{v-2} \\ \cline{1-1}
  \zeta & \\ \cline{1-1} 
\end{array}
= \frac{Q_{b-1}(v+\sum_{j=1}^{b-1}p_{j}+2)Q_{b+1}(v+\sum_{j=1}^{b+1}p_{j}-4)}
{Q_{b-1}(v+\sum_{j=1}^{b-1}p_{j})Q_{b+1}(v+\sum_{j=1}^{b+1}p_{j}-2)}X_{1} 
\end{equation}
where \framebox{$\xi$} and \framebox{$\zeta$} are columns whose  
total length are $a-2$, which do not involve \framebox{$b$} and 
\framebox{$b+1$}; $v=u+h_1$: $h_1$ is some shift parameter; 
the function $X_{1}$ does not contain the function $Q_{b}$.
Obviously, $S_{2}$ is free of color $b$ pole. \\ 
The case $ b \in J_{+}, b+1 \in J_{-} $ : $S_{k} (k\ge 2)$ is the 
summation of the tableaux of the following form  
\begin{eqnarray}
 \begin{array}{|c|l}\cline{1-1}
    \xi & \\ \cline{1-1} 
    b   & _v  \\ \cline{1-1} 
    b+1 & _{v-2} \\ \cline{1-1} 
   \vdots & \\  \cline{1-1} 
    b+1 & _{v-2k+2} \\ \cline{1-1} 
   \zeta & \\ \cline{1-1}
\end{array}
&=& \frac{Q_{b-1}(v+\sum_{j=1}^{b-1}p_{j}+2)
        Q_{b}(v+\sum_{j=1}^{b-1}p_{j}-2k+2)}
       {Q_{b-1}(v+\sum_{j=1}^{b-1}p_{j})
        Q_{b}(v+\sum_{j=1}^{b-1}p_{j}+2)} \label{tabk1} \\ 
&\times & \frac{Q_{b+1}(v+\sum_{j=1}^{b-1}p_{j})}
        {Q_{b+1}(v+\sum_{j=1}^{b-1}p_{j}-2k+2)}X_{2} \nonumber 
\end{eqnarray}
and 
\begin{equation}
 \begin{array}{|c|l}\cline{1-1}
    \xi & \\ \cline{1-1} 
    b+1   & _v \\ \cline{1-1} 
    b+1   & _{v-2} \\ \cline{1-1} 
   \vdots & \\ \cline{1-1} 
    b+1 & _{v-2k+2}\\ \cline{1-1} 
   \zeta & \\ \cline{1-1}
\end{array}
=\frac{Q_{b}(v+\sum_{j=1}^{b-1}p_{j}-2k+1)
       Q_{b+1}(v+\sum_{j=1}^{b-1}p_{j}+2)}
      {Q_{b}(v+\sum_{j=1}^{b-1}p_{j}+1)
       Q_{b+1}(v+\sum_{j=1}^{b-1}p_{j}-2k+2)}X_{2} 
\label{tabk2}
\end{equation}
where \framebox{$\xi$} and \framebox{$\zeta$} are columns with 
total length $a-k$, which do not involve \framebox{$b$} and 
\framebox{$b+1$}; $v=u+h_2$: $h_2$ is some shift parameter; 
the function $X_{2}$ does not contain the function $Q_{b}$.
Obviously, color $b$ residues in  
(\ref{tabk1}) and (\ref{tabk2}) cancel each other
 under the Bethe ansatz equation (\ref{BAE}). \\ 
The case $ b \in J_{-}, b+1 \in J_{+} $ : $S_{k} (k\ge 2)$ is the 
summation of the tableaux of the following form  
\begin{eqnarray}
 \begin{array}{|c|l}\cline{1-1}
    \xi & \\ \cline{1-1} 
    b   & _v  \\ \cline{1-1} 
    \vdots & \\  \cline{1-1}
    b & _{v-2k+4} \\ \cline{1-1}
    b+1 & _{v-2k+2} \\ \cline{1-1} 
   \zeta & \\ \cline{1-1}
\end{array}
&=& \frac{Q_{b-1}(v+\sum_{j=1}^{b-1}p_{j}-2k+2)
        Q_{b}(v+\sum_{j=1}^{b-1}p_{j}+1)}
       {Q_{b-1}(v+\sum_{j=1}^{b-1}p_{j})
        Q_{b}(v+\sum_{j=1}^{b-1}p_{j}-2k+1)} \label{tabk31} \\ 
&\times &  \frac{Q_{b+1}(v+\sum_{j=1}^{b-1}p_{j}-2k)}
        {Q_{b+1}(v+\sum_{j=1}^{b-1}p_{j}-2k+2)}X_{3} \nonumber 
\end{eqnarray}
and 
\begin{equation}
 \begin{array}{|c|l}\cline{1-1}
    \xi & \\ \cline{1-1} 
    b   & _v \\ \cline{1-1} 
    \vdots & \\ \cline{1-1} 
    b   & _{v-2k+4} \\ \cline{1-1} 
    b & _{v-2k+2}\\ \cline{1-1} 
    \zeta & \\ \cline{1-1}
\end{array}
=\frac{Q_{b-1}(v+\sum_{j=1}^{b-1}p_{j}-2k)
       Q_{b}(v+\sum_{j=1}^{b-1}p_{j}+1)}
      {Q_{b-1}(v+\sum_{j=1}^{b-1}p_{j})
       Q_{b}(v+\sum_{j=1}^{b-1}p_{j}-2k+1)}X_{3} 
\label{tabk32}
\end{equation}
where \framebox{$\xi$} and \framebox{$\zeta$} are columns with 
total length $a-k$, which do not involve \framebox{$b$} and 
\framebox{$b+1$}; $v=u+h_3$: $h_3$ is some shift parameter; 
the function $X_{3}$ does not contain the function $Q_{b}$.
Obviously, color $b$ residues in  
(\ref{tabk31}) and (\ref{tabk32}) cancel each other
 under the Bethe ansatz equation (\ref{BAE}). \\ 
The case $b,b+1 \in J_{-}$: $S_{k} (k \ge 2)$ is the summation of 
 the tableaux of the following form 
\begin{eqnarray} 
&& f(k,n,\xi,\zeta,u):= 
\begin{array}{|c|l}\cline{1-1}
    \xi & \\ \cline{1-1} 
    b   & _v \\ \cline{1-1} 
    \vdots & \\ \cline{1-1} 
    b & _{v-2n+2}\\ \cline{1-1} 
    b+1 & _{v-2n} \\ \cline{1-1} 
   \vdots & \\ \cline{1-1} 
    b+1 & _{v-2k+2}\\ \cline{1-1} 
   \zeta & \\ \cline{1-1}
\end{array}  \nonumber \\
&=&\frac{Q_{b-1}(v+\sum_{j=1}^{b-1}p_{j}-2n)Q_{b}(v+\sum_{j=1}^{b-1}p_{j}+1)}
 {Q_{b-1}(v+\sum_{j=1}^{b-1}p_{j})Q_{b}(v+\sum_{j=1}^{b-1}p_{j}-2n+1)} \\ 
&\times & \frac{Q_{b}(v+\sum_{j=1}^{b-1}p_{j}-2k-1)
               Q_{b+1}(v+\sum_{j=1}^{b-1}p_{j}-2n)}
          {Q_{b}(v+\sum_{j=1}^{b-1}p_{j}-2n-1)
          Q_{b+1}(v+\sum_{j=1}^{b-1}p_{j}-2k)} X_{4} 
,\quad  0 \le n \le k \nonumber
\label{tableauxk3}
\end{eqnarray}
where \framebox{$\xi$} and \framebox{$\zeta$} are columns with 
total length $a-k$, which do not involve \framebox{$b$} and 
\framebox{$b+1$}; $v=u+h_4$: $h_4$ is some shift parameter and 
is independent of $n$; the function $X_{4}$ does not have color $b$
 pole and is independent of $n$.
$f(k,n,\xi,\zeta,u)$ has color $b$ poles at
 $u=-h_4-\sum_{j=1}^{b-1}p_{j}+2n-1+u_{l}^{(b)}$
  and $u=-h_4-\sum_{j=1}^{b-1}p_{j}+2n+1+u_{l}^{(b)}$
  for $1 \le n \le k-1$; at 
  $u=-h_4-\sum_{j=1}^{b-1}p_{j}+1+u_{l}^{(b)}$ 
for $n=0$ ; 
at $u=-h_4-\sum_{j=1}^{b-1}p_{j}+2k-1+u_{l}^{(b)}$ for $n=k$. 
Evidently, color $b$ residue at 
$u=-h_4-\sum_{j=1}^{b-1}p_{j}+2n+1+u_{l}^{(b)}$
 in  $f(k,n,\xi,\zeta,u)$ and $f(k,n+1,\xi,\zeta,u)$ 
 for $0\le n \le k-1$ 
 cancel each other under the Bethe ansatz equation (\ref{BAE}). 
 Thus, under the Bethe ansatz equation
  (\ref{BAE}), $\sum_{n=0}^{k}f(k,n,\xi,\zeta,u)$ 
 is free of color $b$ poles, so is $S_{k}$.
\rule{5pt}{10pt} \\ 
Applying Theorem \ref{polefree} to  (\ref{Jacobi-Trudi1}),
 one can show that 
${\cal T}_{\lambda \subset \mu}(u)$ 
is free of poles under the Bethe ansatz equation (\ref{BAE}). 
Thus each term in ${\cal T}_{\lambda \subset \mu}(u)$ has a counterterm 
which cancel the common pole under the Bethe ansatz equation (\ref{BAE}). 
Futhermore the set of all the terms in ${\cal T}_{\lambda \subset \mu}(u)$ 
 forms \symbol{96}Bethe-strap\symbol{39} structure, 
which bears a resemblance to a weight space diagram.  
 See Figure \ref{best} and the relation (\ref{example}) 
 for $\lambda=\phi$ and $\mu=(2^{1},1^{1})$ case; the 
 diagram (\ref{cg}) for $\lambda=\phi$ and $\mu=(1^{1})$ case. 
Consult the references \cite{KS1,S2} for detailed accounts on the 
\symbol{96}Bethe-strap\symbol{39}. 
%%%%%%%%%%%%%%%%%%%%%%%%%%%%%%%%%%%%%%%%%%%%%%%%%%%%%%%%%%%%
\eqreset
\section{Functional equations}
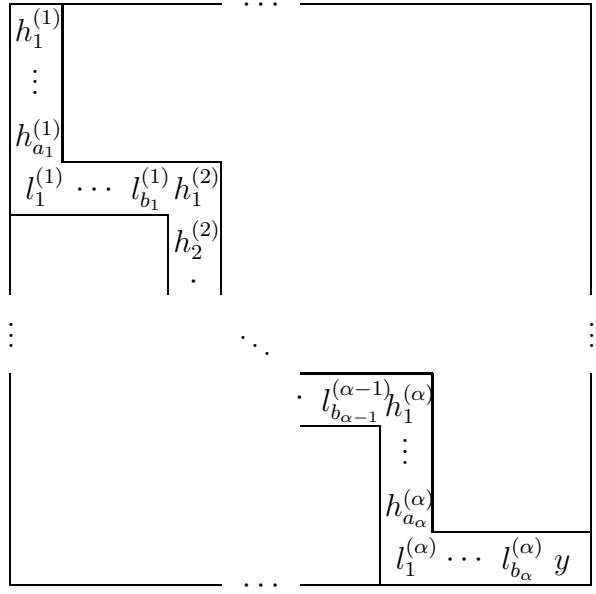
\begin{figure}
  \begin{center}
    \setlength{\unitlength}{2pt}
    \begin{picture}(120,120) 
      \put(0,0){\line(0,1){40}}
      \put(-0.5,45){$\vdots$}
      \put(0,55){\line(0,1){55}}
      \put(10,80){\line(0,1){30}}
      \put(30,55){\line(0,1){15}}
      \put(40,55){\line(0,1){25}}
      \put(70,0){\line(0,1){30}}
      \put(80,10){\line(0,1){30}}
      \put(110,0){\line(0,1){40}}
      \put(109.3,45){$\vdots$}
      \put(110,55){\line(0,1){55}}
      \put(0,0){\line(1,0){40}}
      \put(44,-1.5){$\cdots$}
      \put(55,0){\line(1,0){55}}      
      \put(80,10){\line(1,0){30}}
      \put(55,30){\line(1,0){15}}
      \put(55,40){\line(1,0){25}}
      \put(0,70){\line(1,0){30}}
      \put(10,80){\line(1,0){30}}
      \put(0,110){\line(1,0){40}}
      \put(44,108.5){$\cdots$}
      \put(55,110){\line(1,0){55}}
      \put(1,103){$h_{1}^{(1)}$}
      \put(4,93){$\vdots$}
      \put(1,83){$h_{a_{1}}^{(1)}$}
      \put(3,73){$l_{1}^{(1)}$}
      \put(12.5,73.5){$\cdots$}
      \put(23,73){$l_{b_{1}}^{(1)}$}
      \put(31,73){$h_{1}^{(2)}$}
      \put(31,63){$h_{2}^{(2)}$}
      \put(34,56){$\cdot$}
      \put(43,43){$\ddots$}
      \put(54,34){$\cdot$}
      \put(59,33){$l_{b_{\alpha-1}} ^{(\alpha-1)}$}
      \put(71,32){$h_{1}^{(\alpha)}$}
      \put(74,23){$\vdots$}
      \put(71,13){$h_{a_{\alpha}}^{(\alpha)}$}
      \put(73,3){$l_{1}^{(\alpha)}$}
      \put(82.5,3.5){$\cdots$}
      \put(93,3){$l_{b_{\alpha}}^{(\alpha)}$}
      \put(103,3){$y$}
    \end{picture}
  \end{center}
  \caption{a strip in $(r+2)\times (s+2)$ rectangular subdiagram }
  \label{strip}
\end{figure}

Consider the following Jacobi identity: 
\begin{equation}
  {D}\left[
   \begin{array}{c}
        b \\
        b 
   \end{array}
  \right]
  {D}
   \left[
   \begin{array}{c}
        c \\ 
        c 
   \end{array}
  \right]-
 {D}\left[
   \begin{array}{c}
        b \\
        c 
   \end{array}
  \right]
 {D}\left[
   \begin{array}{c}
        c \\
        b 
   \end{array}
  \right]=
 {D}\left[
   \begin{array}{cc}
        b & c\\
        b & c
   \end{array}
  \right]
   {D},  
   \quad b \ne c    
        \label{jacobi}	
\end{equation} 
where $D$ is the deterement of a matrix and 
${D}\left[
   \begin{array}{ccc}
        a_{1} & a_{2} & \dots \\
        b_{1} & b_{2} & \dots
   \end{array}
   \right]$
is its minor removing $a_{\alpha}$'s rows and 
$b_{\beta}$'s columns.
Set $\lambda = \phi$, $ \mu =(m^a)$ in (\ref{Jacobi-Trudi1}).
 From the relation (\ref{jacobi}), we have
\begin{equation}
  {\cal T}_{m}^{a}(u-1) {\cal T}_{m}^{a}(u+1)  = 
    {\cal T}_{m+1}^{a}(u) {\cal T}_{m-1}^{a}(u)+
    {\cal T}_{m}^{a-1}(u) {\cal T}_{m}^{a+1}(u) 
        \label{t-sys1} 
\end{equation}
where $a,m \ge 1$; ${\cal T}_{m}^{a}(u)={\cal T}_{(m^a)}(u)$: $a,m \ge 1$;
 ${\cal T}_{m}^{0}(u)=1$: $m \ge 0$; ${\cal T}_{0}^{a}(u)=1$: $a \ge 0$.
%
%Solving this functional equation recursively, we can express its solutions 
%in te
The functional equation (\ref{t-sys1}) is a special 
case of Hirota bilinear difference equation \cite{H}. 
For $s=-1$, this functional equation (\ref{t-sys1}) reduces to a 
discretized
 Toda field equation of $A_{r}$ type.
Furthermore, there is a restriction on it,  
which we consider below.
\begin{theorem}\label{vanish} 
 ${\cal T}_{\lambda \subset \mu}(u)=0$ 
if $\lambda \subset \mu$ contains 
a rectangular subdiagram with $r+2$ rows and $s+2$ columns.
\end{theorem}
Proof. Consider a tablau $b$ on this Young superdiagram 
 $\lambda \subset \mu$. Decompose the set $J_{+}$ and $J_{-}$ 
 (\ref{disj}) as  
a union of the disjoint sets:
\begin{eqnarray}
J_{+}=\bigcup_{k=1}^{\alpha} J_{+}^{(k)}: \quad 
J_{+}^{(k)}=\{i_{1}^{(k)},i_{2}^{(k)},\cdots,i_{a_{k}}^{(k)} \}, \\
J_{-}=\bigcup_{k=1}^{\alpha} J_{-}^{(k)}: \quad 
J_{-}^{(k)}=\{j_{1}^{(k)},j_{2}^{(k)},\cdots,j_{b_{k}}^{(k)} \}
\end{eqnarray}
where we assumed, for any $k \in \{ 1,2,\dots ,\alpha \}$, 
\begin{eqnarray}
	i_{\gamma}^{(k)} & = & \sum_{\delta=1}^{k-1}
	 (a_{\delta}+b_{\delta})+\gamma : 
	 \quad \gamma \in \{ 1,2,\dots ,a_k \}, \\
    j_{\gamma}^{(k)} & = & \sum_{\delta=1}^{k-1}
	 (a_{\delta}+b_{\delta})+a_{k}+\gamma :
	 \quad \gamma \in \{ 1,2,\dots ,b_k \} .
\end{eqnarray}
Note that $J_{+}^{(1)}=\phi \quad (a_{1}=0)$, if the minimal
 element in the set $J$ is a member of the set $J_{-}$; 
 $J_{-}^{(\alpha)}=\phi \quad (b_{\alpha}=0)$, if the maximal
 element in the set $J$ is a member of the set $J_{+}$.
 On this rectangular subdiagram, consider a strip, which is a union of 
 $a_{k}\times 1$ rectangular subdiagrams, $1\times b_{k}$ rectangular 
 subdiagrams and $1\times 1$ square subdiagram.
Fill this strip by the elements $\{h_{t}^{(k)}, l_{t}^{(k)} \}$ of $J$ 
so as to meet the admissibility 
conditions (i), (ii) and (iii) (see, Figure \ref{strip}).
For any $k \in \{1,2,\dots \alpha \}$, we find 
\begin{eqnarray}
h_{t}^{(k)} \succeq i_{t}^{(k)} &:& \quad t \in \{1,2,\dots a_{k}\}, \\    
l_{t}^{(k)} \succeq j_{t}^{(k)} &:& \quad t \in \{1,2,\dots b_{k} \}. 
\end{eqnarray}
If $J_{-}^{(\alpha)} \ne \phi \quad (b_{\alpha} \ne 0)$, there is no admissible 
element $y \in J$ since $l_{b_{\alpha}}^{(\alpha)} 
\succeq j_{b_{\alpha}}^{(\alpha)}=r+s+2 \in J_{-}$.
If $J_{-}^{(\alpha)} = \phi \quad (b_{\alpha} = 0)$, there is no admissible 
element $y \in J$ since $h_{a_{\alpha}}^{(\alpha)} 
\succeq i_{a_{\alpha}}^{(\alpha)}=r+s+2 \in J_{+}$. 
Then there is no admissible tableau on this Young superdiagram.
  \rule{5pt}{10pt} \\
Remark:  The spectrum of fusion model was discussed 
in references \cite{DM,MR}
from the point of view of representation theory 
and the corresponding theorem 
%for the grading $p_{i}=1:1 \le i \le r+1; 
%p_{i}=-1:r+2 \le i \le r+s+2 $ 
 was discussed. \\
As a corollary, we have 
\begin{equation}
{\cal T}_{m}^{a}(u)=0 \quad {\rm for} \quad 
a \ge r+2  \quad {\rm and} \quad  m \ge s+2. \label{vanish2}
\end{equation}  
Applying the relation (\ref{vanish2}) to (\ref{t-sys1}),
 we obtain 
\begin{equation}
  {\cal T}_{m}^{r+1}(u-1) {\cal T}_{m}^{r+1}(u+1)  = 
    {\cal T}_{m+1}^{r+1}(u) {\cal T}_{m-1}^{r+1}(u)
     \quad m \ge s+2,
        \label{laplace1} 
\end{equation}
\begin{equation}
  {\cal T}_{s+1}^{a}(u-1) {\cal T}_{s+1}^{a}(u+1)  = 
    {\cal T}_{s+1}^{a-1}(u) {\cal T}_{s+1}^{a+1}(u)
     \quad a \ge r+2.
        \label{laplace2} 
\end{equation}
%%%%%%%%%%%%%%%%%%%%%%%%%%%%%%%%%%%%%%%%%%%%%%%%%%%%%%%%%%%%%%%%%%%%%
\eqreset
\section{On the equivalence of the Bethe ansatz equations}
Bares et. al. \cite{BCFH} showed that Lai's \cite{L}
 representation of the Bethe ansatz 
equation on the supersymmetric $t-J$ model is equivalent to 
Sutherland 's one \cite{Su} under the particle-hole transformation. 
Moreover, following reference \cite{BCFH}, Essler and Korepin \cite{EK} 
showed that Sutherland 's \cite{Su} representation of the Bethe ansatz 
equation on the supersymmetric $t-J$ model is equivalent to 
 the one originate from the grading $(p_{1},p_{2},p_{3})=(-1,1,-1)$
for Lie superalgebra $sl(1|2)$. Then the eqivalence 
 of three different sets of Bethe ansatz equations on the 
 supersymmetric $t-J$ model, which originate from   
$\frac{3!}{1! 2!}=3$ different gradings for $sl(1|2)$
 was established.
 Futhermore Essler et. al. \cite{EKS2} established the equivalence of six  
different sets of Bethe ansatz equations on the supersymmetric 
extended Hubbard model, 
 which originate from $\frac{4!}{2! 2!}=6$ different gradings  
for Lie superalgebra $sl(2|2)$ (see, Figure \ref{dynkin}).
 Now, following reference \cite{BCFH},
  we discuss relations among the sets of Bethe 
ansatz equations (\ref{BAE}) 
for different $\frac{(r+s+2)!}{(r+1)!(s+1)!}$ 
gradings $\{p_{j}\}$ (\ref{grading})
 or different sets of simple root systems 
inequivalent under the Weyl group ${\cal W}({\cal G})$
of Lie superalgebra $sl(r+1|s+1)$.

In this section, we assume that $q=1$. 
For some $b$ ($2 \le b \le r+s $), we assume $p_{b}p_{b+1}=-1$. 
Namely, $b$ th simple root $\alpha_{b}$
 is an odd root with $(\alpha_{b}|\alpha_{b})=0$.  In this case, 
$b$ th Bethe ansatz equation in (\ref{BAE}) has the following form   
\begin{equation}
 1=\frac{Q_{b-1}(u_{k}^{(b)}-p_{b}) Q_{b+1}(u_{k}^{(b)}-p_{b+1})}
 {Q_{b-1}(u_{k}^{(b)}+p_{b}) Q_{b+1}(u_{k}^{(b)}+p_{b+1})},\quad 
    k=1,2,\dots,N_{b}.\label{bae2}
\end{equation}
Define the polynomial 
\begin{equation}
 f(z)=Q_{b-1}(z+p_{b}) Q_{b+1}(z+p_{b+1})
-Q_{b-1}(z-p_{b}) Q_{b+1}(z-p_{b+1}).\label{ffun}
\end{equation}
Among the roots of the equation $f(z)=0$, $N_{b}$ of which are 
$\{u_{k}^{(b)}\}_{1 \le k \le N_{b}}$. 
So $\{f(u_{k}^{(b)})=0\}_{1 \le k \le N_{b}}$ 
reproduces the Bethe ansatz equation (\ref{bae2}).
Let the rest of the roots be $\{\tilde{u}_{k}^{(b)}\}_{
1 \le k \le \tilde{N}_{b}}$. Then $\{f(\tilde{u}_{k}^{(b)})=0\}_{
1 \le k \le \tilde{N}_{b}}$ reduces to  
the Bethe ansatz equation of the form 
\begin{equation}
 1=\frac{Q_{b-1}(\tilde{u}_{k}^{(b)}+p_{b}) 
    Q_{b+1}(\tilde{u}_{k}^{(b)}+p_{b+1})}
   {Q_{b-1}(\tilde{u}_{k}^{(b)}-p_{b}) 
   Q_{b+1}(\tilde{u}_{k}^{(b)}-p_{b+1})},\quad 
    k=1,2,\dots,\tilde{N}_{b}. \label{baet2}
\end{equation}
Thanks to the residue theorem, the following relation holds 
\begin{eqnarray}
&& \sum_{j=1}^{{N}_{b}}\frac{1}{2\pi i} \int_{C_{j}}dz \frac{1}{i}
 {\bf Log} 
 \frac{z-u_{l}^{(b-1)}-p_{b}}{z-u_{l}^{(b-1)}+p_{b}} \  
 \frac{d}{dz} {\bf Log} f(z) \nonumber \\ 
& =& \sum_{j=1}^{{N}_{b}} \frac{1}{i} {\bf Log} 
 \frac{u_{j}^{(b)}-u_{l}^{(b-1)}-p_{b}}
      {u_{j}^{(b)}-u_{l}^{(b-1)}+p_{b}} \label{log1} 
\end{eqnarray}
where $C_{j}$ denotes contor around $u_{j}^{(b)}$. 
We assume the branch cut of the logarithm in the 
left hand side of (\ref{log1}) 
extends from $u_{l}^{(b-1)}-p_{b}$ to $u_{l}^{(b-1)}+p_{b}$.     
The left hand side of 
the relation (\ref{log1}) can be rewritten as follows 
\begin{eqnarray}
 -\sum_{j=1}^{{\tilde{N}}_{b}} \frac{1}{i} {\bf Log} 
 \frac{\tilde{u}_{j}^{(b)}-u_{l}^{(b-1)}-p_{b}}
      {\tilde{u}_{j}^{(b)}-u_{l}^{(b-1)}+p_{b}} 
     +\frac{1}{i} {\bf Log} 
     \frac{f(u_{l}^{(b-1)}+p_{b})}{f(u_{l}^{(b-1)}-p_{b})}.
\end{eqnarray}
Then the following relation holds 
\begin{equation}
 -1=\frac{Q_{b-1}(u_{l}^{(b-1)}+2p_{b}) 
    \tilde{Q}_{b}(u_{l}^{(b-1)}-p_{b})
    Q_{b}(u_{l}^{(b-1)}-p_{b})}
    {Q_{b-1}(u_{l}^{(b-1)}-2p_{b}) 
    \tilde{Q}_{b}(u_{l}^{(b-1)}+p_{b})
    Q_{b}(u_{l}^{(b-1)}+p_{b})},\quad 
    l=1,2,\dots,N_{b-1}. \label{baem1}
\end{equation}
where $\tilde{Q}_{b}(u)=
\prod_{j=1}^{\tilde{N}_{b}}(u-\tilde{u}_{j}^{(b)})$.
Noting that the relation 
\begin{equation}
 \frac{Q_{b-1}(u_{l}^{(b-1)}+p_{b-1}+p_{b})}
      {Q_{b-1}(u_{l}^{(b-1)}-p_{b-1}-p_{b})}
 =\frac{Q_{b-1}(u_{l}^{(b-1)}+2p_{b})}{Q_{b-1}(u_{l}^{(b-1)}-2p_{b-1})},
\end{equation}
we find that the $b-1$th Bethe ansatz equation in 
(\ref{BAE}) has the form:
\begin{eqnarray}
 -1=(-1)^{{\rm deg}(\alpha_{b-1})}
       \frac{Q_{b-2}(u_{l}^{(b-1)}-p_{b-1})
             Q_{b-1}(u_{l}^{(b-1)}+2p_{b}) 
             Q_{b}(u_{l}^{(b-1)}-p_{b})}
            {Q_{b-2}(u_{l}^{(b-1)}+p_{b-1})
             Q_{b-1}(u_{l}^{(b-1)}-2p_{b-1})
             Q_{b}(u_{l}^{(b-1)}+p_{b})}, \label{bae1} \\
    l=1,2,\dots,N_{b-1}.\nonumber
\end{eqnarray}
Combining these two equations (\ref{bae1}) and (\ref{baem1}), we obtain 
\begin{eqnarray}
 -1=(-1)^{{\rm deg}(\tilde{\alpha}_{b-1})}
\frac{Q_{b-2}(u_{l}^{(b-1)}-p_{b-1})
             Q_{b-1}(u_{l}^{(b-1)}-2p_{b}) 
             \tilde{Q}_{b}(u_{l}^{(b-1)}+p_{b})}
            {Q_{b-2}(u_{l}^{(b-1)}+p_{b-1})
             Q_{b-1}(u_{l}^{(b-1)}-2p_{b-1})
             \tilde{Q}_{b}(u_{l}^{(b-1)}-p_{b})}, \label{baet1} \\ 
    l=1,2,\dots,N_{b-1} \nonumber
\end{eqnarray}
where ${\rm deg}(\tilde{\alpha}_{b-1})={\rm deg}(\alpha_{b-1})+1 
\quad mod \> 2$.
The following relation is valid 
\begin{eqnarray}
&& \sum_{j=1}^{{N}_{b}}\frac{1}{2\pi i} \int_{C_{j}}dz \frac{1}{i}
 {\bf Log} 
 \frac{z-u_{l}^{(b+1)}-p_{b+1}}{z-u_{l}^{(b+1)}+p_{b+1}} \  
 \frac{d}{dz} {\bf Log} f(z) \nonumber \\ 
&=& \sum_{j=1}^{{N}_{b}} \frac{1}{i} {\bf Log} 
 \frac{u_{j}^{(b)}-u_{l}^{(b+1)}-p_{b+1}}
      {u_{j}^{(b)}-u_{l}^{(b+1)}+p_{b+1}} \label{log2} \\ 
&=& -\sum_{j=1}^{{\tilde{N}}_{b}} \frac{1}{i} {\bf Log} 
 \frac{\tilde{u}_{j}^{(b)}-u_{l}^{(b+1)}-p_{b+1}}
      {\tilde{u}_{j}^{(b)}-u_{l}^{(b+1)}+p_{b+1}} 
     +\frac{1}{i}
      {\bf Log} \frac{f(u_{l}^{(b+1)}+p_{b+1})}{f(u_{l}^{(b+1)}-p_{b+1})} 
     \nonumber 
\end{eqnarray}
where $C_{j}$ denotes contor around $u_{j}^{(b)}$. 
We assume the branch cut of the logarithm in left hand side of (\ref{log2}) 
extends from $u_{l}^{(b+1)}-p_{b+1}$ to $u_{l}^{(b+1)}+p_{b+1}$.   
This equation reduces to the following equation:
\begin{eqnarray}
 -1=\frac{Q_{b+1}(u_{l}^{(b+1)}+2p_{b+1}) 
    \tilde{Q}_{b}(u_{l}^{(b+1)}-p_{b+1})
    Q_{b}(u_{l}^{(b+1)}-p_{b+1})}
    {Q_{b+1}(u_{l}^{(b+1)}-2p_{b+1}) 
    \tilde{Q}_{b}(u_{l}^{(b+1)}+p_{b+1})
    Q_{b}(u_{l}^{(b+1)}+p_{b+1})}, \label{baem2} \\
    l=1,2,\dots,N_{b+1}. \nonumber 
\end{eqnarray}
$b+1$th Bethe ansatz equation in (\ref{BAE}) has the form:
\begin{eqnarray}
 -1=(-1)^{{\rm deg}(\alpha_{b+1})}
       \frac{Q_{b}(u_{l}^{(b+1)}-p_{b+1})
             Q_{b+1}(u_{l}^{(b+1)}+2p_{b+2}) 
             Q_{b+2}(u_{l}^{(b+1)}-p_{b+2})}
            {Q_{b}(u_{l}^{(b+1)}+p_{b+1})
             Q_{b+1}(u_{l}^{(b+1)}-2p_{b+1})
             Q_{b+2}(u_{l}^{(b+1)}+p_{b+2})}, \label{bae3} \\
    l=1,2,\dots,N_{b+1}.\nonumber
\end{eqnarray}
Combining these two equations (\ref{baem2}) and (\ref{bae3}), we obtain 
\begin{eqnarray}
 -1=(-1)^{{\rm deg}(\tilde{\alpha}_{b+1})}
 \frac{\tilde{Q}_{b}(u_{l}^{(b+1)}+p_{b+1})
             Q_{b+1}(u_{l}^{(b+1)}+2p_{b+2}) 
             Q_{b+2}(u_{l}^{(b+1)}-p_{b+2})}
            {\tilde{Q}_{b}(u_{l}^{(b+1)}-p_{b+1})
             Q_{b+1}(u_{l}^{(b+1)}+2p_{b+1})
             Q_{b+2}(u_{l}^{(b+1)}+p_{b+2})}, \label{baet3} \\ 
    l=1,2,\dots,N_{b+1}.\nonumber
\end{eqnarray}
where ${\rm deg}(\tilde{\alpha}_{b+1})={\rm deg}(\alpha_{b+1})+1 
\quad mod \> 2$.
Note that the set of the equations
 (\ref{bae1}),  (\ref{bae2}) and (\ref{bae3}) 
 is transfered to the equivalent set of 
the equations (\ref{baet1}), (\ref{baet2}) and (\ref{baet3}) under the 
following transformation: 
\begin{equation}
 (p_{b},p_{b+1},N_{b},\{u_{k}^{(b)}\}) \longrightarrow 
 (-p_{b},-p_{b+1},\tilde{N}_{b},\{\tilde{u}_{k}^{(b)}\}). \label{trans} 
\end{equation}
One can also develop similar argument for $p_{1}p_{2}=-1$ case using 
the function 
\begin{equation}
f(z)=P_{1}(z+p_{1}) Q_{2}(z+p_{2})
-P_{1}(z-p_{1}) Q_{2}(z-p_{2})
\end{equation}
instead of the function (\ref{ffun}) and for
 $p_{r+s+1}p_{r+s+2}=-1$ case using the function 
\begin{equation}
f(z)=Q_{r+s}(z+p_{r+s+1})-Q_{r+s}(z-p_{r+s+1})
\end{equation}
instead of the function (\ref{ffun}).
 Then the set of the Bethe ansatz equations (\ref{BAE}) is 
transfered to the equivalent set of the Bethe ansatz equations under the  
transformation (\ref{trans}) for $p_{b}p_{b+1}=-1$. 
Therefore, taking notice of a change of the grading
 $\{p_{j}\}$ or odd simple root
 $\alpha_{b}$ with $(\alpha_{b}|\alpha_{b})=0$ 
 and applying the transformation (\ref{trans}) repeatedly to the set of the
 Bethe ansatz equations (\ref{BAE}) with any one of the grading $\{p_{j}\}$ 
 for $sl(r+1|s+1)$, one can get the set of the
 Bethe ansatz equations with any other grading $\{p_{j}\}$ 
 for $sl(r+1|s+1)$. 
  Furthermore, we note that the transformation (\ref{trans}) corresponds 
   to the reflection $\omega_{b} \in {\cal SW}({\cal G})$ 
   for odd simple root $\alpha_{b}$ 
with $(\alpha_{b}|\alpha_{b})=0$.
 This fact follows from the relations: 
 $(\omega_{\alpha_{b}}(\alpha_{b})|\omega_{\alpha_{b}}(\alpha_{b+1}))=
 -(-p_{b+1})$, 
 $(\omega_{\alpha_{b}}(\alpha_{b-1})|\omega_{\alpha_{b}}(\alpha_{b-1}))=
 p_{b-1}+(-p_{b})$,  etc.
%%%%%%%%%%%%%%%%%%%%%%%%%%%%%%%%%%%%%%%%
\eqreset
\section{Summary and discussion}
In the present paper, we have executed analytic Bethe ansatz 
 based upon the 
Bethe ansatz equations (\ref{BAE}) with any simple root systems of 
 the Lie superalgebra $sl(r+1|s+1)$. 
 Pole-freeness of eigenvalue formula of transfer matrix in 
 dressed vacuum form was shown for a 
wide class of finite dimensional representations labeled by
 skew-Young superdiagrams.
 Functional relation has been given especially for the eigenvalue formulae 
 of transfer matrices in dressed vacuum form 
 labeled by rectangular Young superdiagrams, 
 which is a special case of Hirota bilinear difference equation with  
 a restrictive relation. There are earlier results
  \cite{T2} for the 
  distinguished simple root system of  
  $sl(r+1|s+1)$, many of which are special case of 
  the results in the present paper. 
  We discussed how the set of the Bethe ansatz equations 
  for any simple root system of 
   $sl(r+1|s+1)$ is related to the one for any other simple 
  root system of $sl(r+1|s+1)$ under the particle-hole 
  transformation. And then, we pointed 
 out that the particle-hole transformation is connected with the 
 reflection with respect to the element of the Weyl supergroup for 
 odd simple root $\alpha $ with $(\alpha | \alpha)=0$.
 
 It should be emphasized that our method explained in the present paper
 is still valid even if such factors like gauge factor, extra sign 
(different from $(-1)^{{\rm deg}(\alpha_{a})}$ in (\ref{BAE})), 
 etc. appear in the Bethe ansatz equation (\ref{BAE}) 
as long as such factors do not influence the analytical property of
 the right hand side of the Bethe ansatz equation (\ref{BAE}). 
  
  In reference \cite{LWZ}, functional relations for any fusion 
 type transfer matrices associated with any (not always rectangular) 
 Young diagrams of simple Lie algebra $A_{r}$ was given. 
 Similar functional relations for suitable boundary conditions will be 
 also valid for  $sl(r+1|s+1)$ case. 

 In reference \cite{FR}, coincidence between 
 the free field realization of the generators 
 of $U_{q}({\cal G}^{(1)})$ and eigenvalue formulae \cite{KS1}  
 of transfer matrices in dressed vacuum form in the analytic 
 Bethe ansatz was discussed associated 
 with classial simple Lie algebras ${\cal G}$.
 As for a Lie superalgebra ${\cal G}$ case, nobody has studied 
 such a relation so far. A deeper inspection will 
 be desirable.
  
It will be interesting problems to extend a similar analysis  
discussed in this paper for other Lie superalgebras, such 
as $B(m|n),C(n)$ and $D(m|n)$.

Finally we note that functional relations among fusion transfer matrices at 
{\it finite} temperatures have been given in
 the preprint \cite{JKS} quite recently
 using quantum transfer matrix approach.
  In addition, these functional relations are transformed into 
 TBA equations without using string hypothesis. 
\\
{\bf Acknowledgments} \\  
The author would like to thank Professor A. Kuniba for  
encouragement. He also thanks Dr J. Suzuki for discussions. 

\end{document}